\documentclass[final,5p,times,twocolumn]{elsarticle}

\usepackage{amssymb}
\usepackage{gensymb}
\usepackage[T1]{fontenc}
\usepackage{amsmath}
\usepackage{xcolor, soul}
\usepackage{subcaption}

\journal{Nuclear Instruments and Methods in Physics Research A}

\begin{document}

\begin{frontmatter}

\title{Experimental validation of the Gaseous Proton Recoil Telescope for quasi-absolute neutron flux measurements}

\author[inst1,inst2]{Carole Chatel\corref{cor1}\fnref{cor2}}\ead{c.chatel@gsi.de}
\cortext[cor1]{Principal corresponding author:}
\fntext[cor2]{Present address: GSI Helmholtzzentrum für Schwerionenforschung, 64291, Darmstadt, Germany}
\author[inst2]{Ludovic Mathieu}
\author[inst2]{Mourad Aïche}
\author[inst2]{Abdel Rebii}
\author[inst2]{Hedi El Amami}
\author[inst2]{Benoît Dufort}
\author[inst1,inst3]{Maria Diakaki}
\author[inst1]{Olivier Bouland$\dagger$}
\author[inst1]{Gilles Noguère}

\affiliation[inst1]{ 
            addressline={CEA, DES/IRESNE/DER/SPRC/LEPh}, 
            city={Saint Paul Lez Durance},
            postcode={13008},
            country={France}}

\affiliation[inst2]{
            addressline={LP2i, UMR5797, Universit\'e de Bordeaux, CNRS}, 
            city={Gradignan},
            postcode={F-33170}, 
            country={France}}

\affiliation[inst3]{
            addressline={National Technical University of Athens, Department of Physics, Zografou Campus}, 
            city={Athens},
            country={Greece}}

\begin{abstract}
The accuracy of neutronics simulations of actual or future reactor cores is nowadays driven by the precision of the nuclear data used as input. Among the most important neutron-induced fission cross sections to understand well are the actinides. It is, indeed, of primary importance to know accurately these cross sections around 1 MeV for the safety of Generation IV reactors. High accuracy measurements of neutron flux are essential for accurate cross section measurements; measurements of this flux with respect to the $^1$H(n,n)p cross section can be made with the proton recoil technique. For an accurate measurement below 1 MeV, the Gaseous Proton Recoil Telescope (GPRT) is developed and characterized, with the aim to provide quasi-absolute neutron flux measurements with an accuracy better than 2$\%$. This detector is composed of a double ionization chamber with a Micromegas segmented detection plane. The pressure of the gas can be adjusted to protons stopping range – and therefore to neutrons energy. An accurate neutron flux measurement requires that the GPRT has an intrinsic efficiency of 100\%, and thus an important effort has been made to verify this. An alpha source and proton micro-beam have been used and the intrinsic efficiency is confirmed to be 100\%. Additionally, the dead-time of the detector has been investigated on a test bench, and is found to be 7.3~ms.
\end{abstract}

\begin{keyword}
Proton recoil \sep neutron flux \sep cross section measurements  \sep Micromegas \sep Time Projection Chamber
\end{keyword}

\end{frontmatter}

\section{Introduction}
\label{Introduction}
The development of fast neutron reactors (Generation IV) \cite{HPRL} requires an accurate measurement of the neutron flux. The accuracy of neutronics calculation codes for reactor cores are nowadays primarily determined by the accuracy of the nuclear evaluations provided as inputs. Among the required improvements are the knowledge of cross sections of neutron-induced reactions, particularly on heavy nuclei. However, for most actinide fission cross sections, discrepancies between different measurements or between different evaluations around 1 MeV remain, despite a general effort from the nuclear community. Generation IV reactors have maximum fission rates between several hundreds keV and a few MeV of neutron energy. It is therefore of prior importance to reduce the uncertainties of cross sections in this neutron energy range for the nuclei of interest \cite{HPRL}. Such cross section measurements require measuring the used neutron flux very accurately, which is generally achieved by measuring the cross section with respect to a secondary standard reaction such as $^{235}$U(n,f) or $^{238}$U(n,f) (see \cite{ChatelPhD} ch. 4.3 for the $^{242}$Pu). However, referencing $^{235}$U and $^{238}$U is quite complex as the data are evolving and sometimes even contradictory. For example for the $^{235}$U between 200~keV and 20~MeV, the uncertainties are given to be between 0.6\% and 1.1\% in \cite{AIEA} in 2007, while between 1.3\% and 2.0\% in \cite{Carlson} in 2018. A similar situation is observed for $^{238}$U for energies between 1~MeV and 20~MeV, with uncertainties from 0.6\% to 1.3\% in \cite{AIEA} and between 1.3\% and 1.6\% in \cite{Carlson}. In addition, differences between databases remain small for $^{235}$U (maximum 1.3\% at 800~keV), but are dramatic for $^{238}$U (up to 45\% difference at 1.16~MeV with the new ENDF/B-VIII.0 version \cite{ENDF} compared to ENDF/B-VII.1 \cite{ENDF} or JEFF3.3 \cite{JEFF}).

\begin{figure}[h!]
	\centering
	\includegraphics[width=0.45\linewidth]{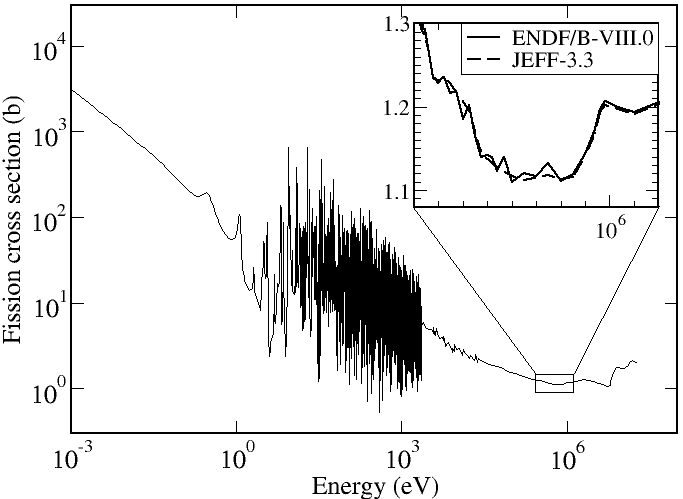}
	\caption{$^{235}$U fission cross section. Data from the JEFF3.3 and ENDF/B-VIII evaluations \cite{JEFF,ENDF}.}
	\label{235U-XS}
\end{figure}

Moreover, using the same standard for different measurements creates a strong correlation between those measurements. Additionally, the measurement of structures present in a fission cross section will be correlated to those present in the $^{235}$U(n,f) cross section, as one can see in figure \ref{235U-XS}.

For this reason, it is essential to carry out independent and accurate measurements using a different standard. A primary standard reaction is preferable, (reason: e.g. to avoid correlations present when using a secondary standard). With an accuracy around 0.4\% between 200~keV and 2.6~MeV \cite{AIEA,Carlson}, a strong cohesion between databases and reproducible with \textit{ab initio} calculations, the $^{1}$H(n,n)p elastic scattering cross section is well suited for this purpose. Moreover, it is structureless, inducing no dependence between the measured structures and the ones from the reference, unlike the $^{235}$U(n,f) cross section. 

Choosing the $^{1}$H(n,n)p cross section as reference requires the usage of the proton recoil technique, which consists of converting the neutron flux into a proton flux that is significantly easier to detect.  For a neutron energy range between 1 and 70~MeV, silicon junctions are appropriate for this purpose \cite{Paola2016,Paola2017,Si1,Si2}. 
For cross section measurements with mono-energetic neutrons, the setup must be placed as close as possible to the neutron source to maximize the neutron flux. This means that many $\gamma$-rays and electrons, produced by the neutron source, are detected in the recoil-proton detector \cite{Paola2016}. This background signal is huge at low energies and prevents the number of recoil protons from being accurately counted. One possibility to overcome this drawback is to adapt the silicon detector thickness to reduce its sensitivity to the electron and $\gamma$-ray background. This solution, however, is too restrictive when working with radioactive targets and not efficient enough for neutrons of a few hundred~keV \cite{Paola2017}. 

Other proton recoil detectors exist, but each come with their own limitations. Plastic scintillators \cite{Plast1,Plast2,Plast3,Plast4,Plast5} must work with a high detection threshold to discriminate neutrons from $\gamma$-rays. Moreover, the efficiency of plastic scintillators is difficult to obtain accurately, requires simulations to account for the Compton effects from $\gamma$-rays. Proportional counters \cite{Counter1,Counter2,Counter3} can also be used for proton recoil techniques, however the efficiency of this kind of detectors must be reconstructed by simulation. This implies that a precise neutron flux measurement relies mostly on complex simulations \cite{Counter1,CountSimu1,CountSimu2}, which is undesirable for an accurate cross section measurement.

Thus, currently available detectors do not meet all the needs for an independent fission cross section measurement. To this end, the Aval du Cycle et \'Energie Nucl\'eaire team of the Laboratoire de Physique des 2 infinis (LP2i) has therefore decided to develop the Gaseous Proton Recoil Telescope (GPRT) \cite{ChatelPhD,ChatelAnimma,ChatelND,Paola2018,Paola2019} in order to accurately measure neutron fluxes for neutron energies between 0.2~MeV and 2.5~MeV. The prototype detector and its acquisition system are presented and its intrinsic efficiency is discussed.

\section{GPRT description}
\label{description}

\subsection{General description}
\label{general}

\begin{figure}[h!]
	\centering
\begin{subfigure}{1\linewidth}
	\centering
	\includegraphics[width=1\linewidth]{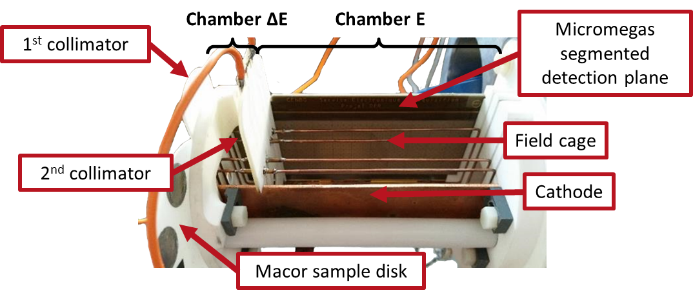}
	\caption{\label{GPRT-Photo}}
\end{subfigure}

\begin{subfigure}{1\linewidth}
	\centering
	\includegraphics[width=0.55\linewidth]{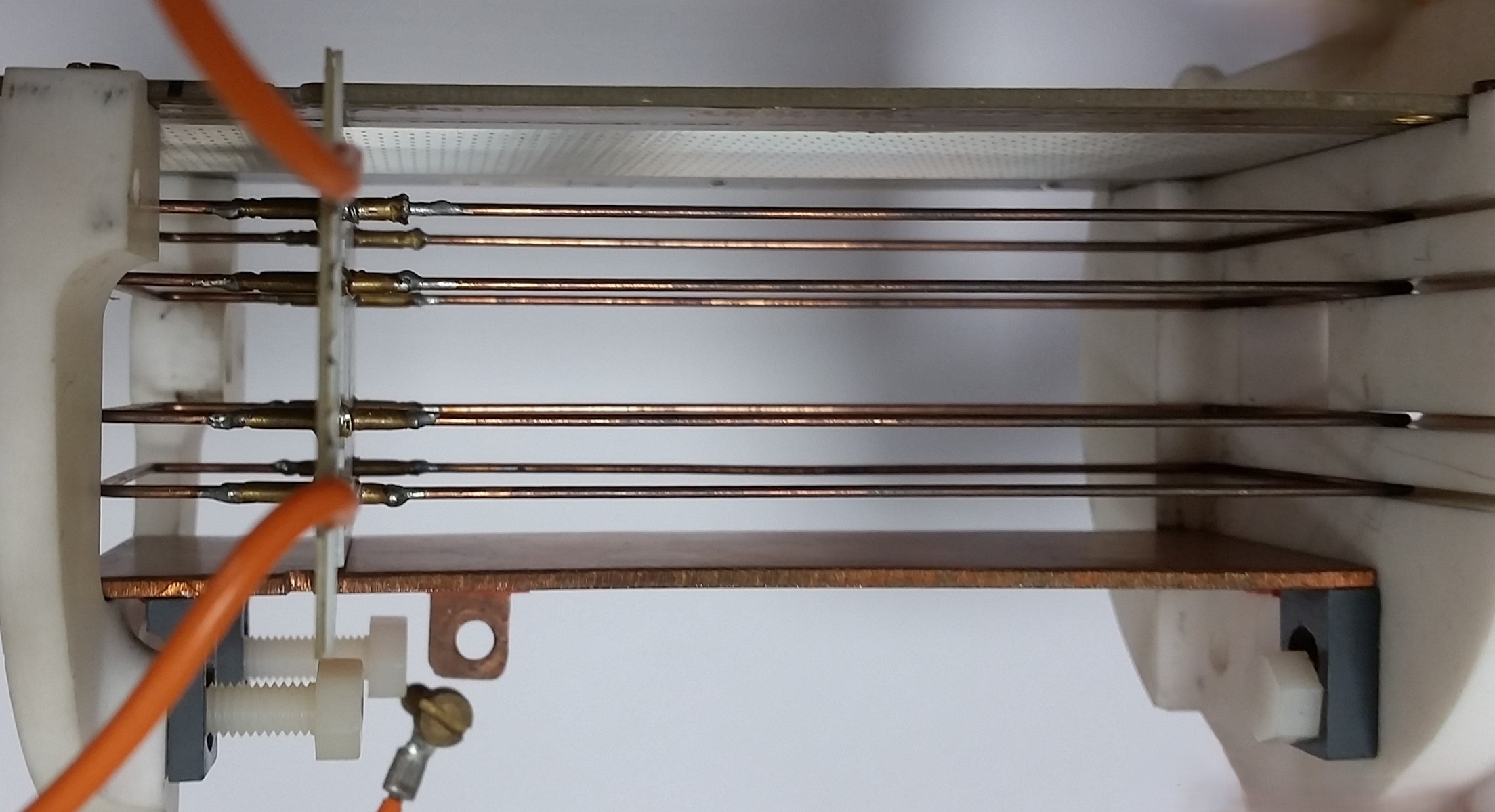}
	\caption{\label{GPRT-above}}
\end{subfigure}
	\caption{\label{GPRT_Photo} Pictures of the Gaseous Proton Recoil Telescope (a) with legend; (b) view from above.}
\end{figure}

\begin{figure}[h!]
	\centering
	\includegraphics[width=0.8\linewidth]{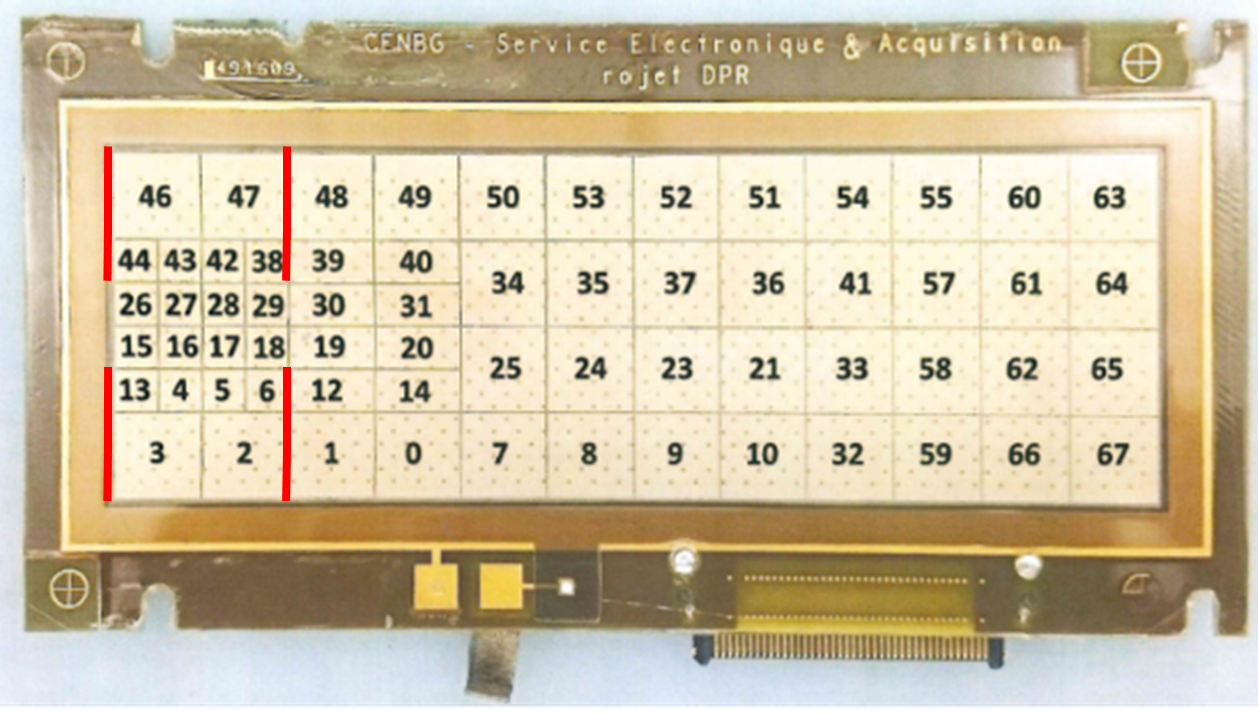}
	\caption{Picture of the Micromegas detector plane. Red thick lines represent the position and aperture of both collimators}
	\label{GPRT_Plane}
\end{figure}

The GPRT is composed of a 12~cm long $\Delta$E-E double ionization chamber. A 10 mm diameter collimator is located at the entrance of the detector, and a second identical collimator is situated 2~cm away from the first, separating the two ionization chambers. Two electrodes, separated by 4 cm of gas, define the geometry of the sides and provide an electric field. In the ionization chambers, the primary electrons generated by the passage of a charged particle drift towards the detection plane. This plane uses the Micromegas technology \cite{Micromegas} to amplify the signal. It consists of an anode plane (12x4~cm$^{2}$), segmented in 64 pads to reconstruct the track of the particle \cite{ChatelPhD,ChatelAnimma,ChatelND}, and an electrified mesh of the same dimensions, 125~$\mu$m above the segmented plane. A field cage is used to make the electric field uniform between the cathode and the mesh. Electrostatic simulations have been realized to verify the homogeneity of this field \cite{Paola2018}. The aim of the first collimator is to define properly the surface of the H-rich sample that will eject recoil-protons in the detector. The purpose of the second collimator is to create a physical and precise delimitation between the $\Delta$E-chamber and the E-chamber, allowing to define accurately the geometric efficiency of the detection system. This also allows to limit the energy of the scattered protons detected in the E-chamber, whose energy is close to the one of the incident neutrons. The segmentation of the detection plane aims to reconstruct the particle's track to verify that it went through both collimators. It has then been chosen that the pads around the collimators would be smaller than in the rest of the detectors, while remaining in the limit of 64 electronic channels (0.5*0.5~cm$^{2}$ before the second collimator, 0.5*1~cm$^{2}$ after the second collimator, 1*1~cm$^{2}$ in the rest of the detector). In front of the entrance of the detector, a holder is placed  (named "Macor sample disk"), on which several H-rich foils of different thicknesses can be mounted in front of the detector. Pictures of the setup are shown in figure~\ref{GPRT_Photo} and the segmented detection plane is enhanced in figure~\ref{GPRT_Plane}.

To perform an accurate measurement, it is essential that as few recoil protons as possible enter the GPRT from sources other than the H-rich sample. Consequently, an H-free gas has been chosen: a mixture of 70\% N$_{2}$ and 30\% CO$_{2}$. The gas pressure can be adapted such that protons are stopped within the geometry of the detector. The structures are also made of H-free materials, metal or macor (white materials in figure~\ref{GPRT_Photo}). The H-rich foil thickness is adapted to the neutron energy, in such a way as to limit the energy loss of the recoil proton in the foil as much as possible. The electric field must then be adapted to the chosen pressure of the gas by adjusting the polarization voltages of the cathode, field cage and mesh. The bias voltages must be as high as possible to maximize the amplification gain but low enough to avoid electrical breakdowns. 

\subsection{Acquisition system}
\label{Acquisition}

The GPRT acquisition system uses the Single AGET Module (SAM) acquisition, derived from the General Electronic for TPC (GET) acquisition \cite{GET} and designed in the LP2i laboratory. Its principle is as follows:

Each pad of the Micromegas detection plane collects charges and sends signals to three successive cards that will process them. The Zap card is the first one. Its main goal is to protect the two other cards from voltage spikes. 

The second card is the SAM card. Based on the use of the AGET (Asic for GET) chip from the CEA/IRFU, Saclay, it is mainly used to preamplify the signal and to manage the trigger. The signal is sent to a discriminator that compares it to the conditions chosen by the user: the amplitude and multiplicity thresholds. Thus, in order to be kept for subsequent analysis, signals must have (a) a minimum number of pads activated that have been designated as triggers; (b) those pads must have a signal amplitude higher than the amplitude threshold.

Finally, the third card is the ZedBoard, a commercial card from Xilinx based on Field Programmable Gate Array (FPGA) type Zynq7020. Its firmware allows to reduce the size of the data and to transfer them - event and timing - to the computer, with a clock of 25 MHz. A channel (pad) has 512 points and each of them requires 40~ns to be read. The dead time is therefore theoretically 20.5~$\mu$s per channel, or 1.4~ms for the 64 channels. This number will be discussed in section~\ref{efficiency}.

\begin{figure}[h!]
	\centering
	\includegraphics[width=0.8\linewidth]{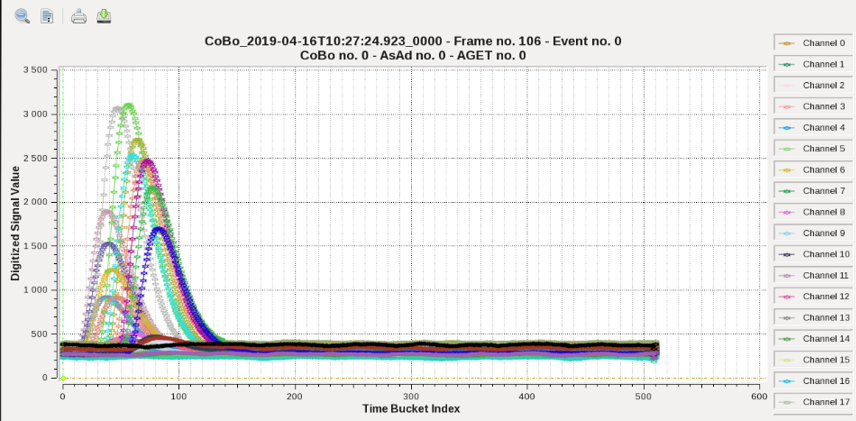}
	\caption{Example of a recorded spectrum obtained with GetController with an $\alpha$ source}
	\label{Cobo-alpha}
\end{figure}

\begin{figure}[h!]
	\centering
	\includegraphics[width=0.8\linewidth]{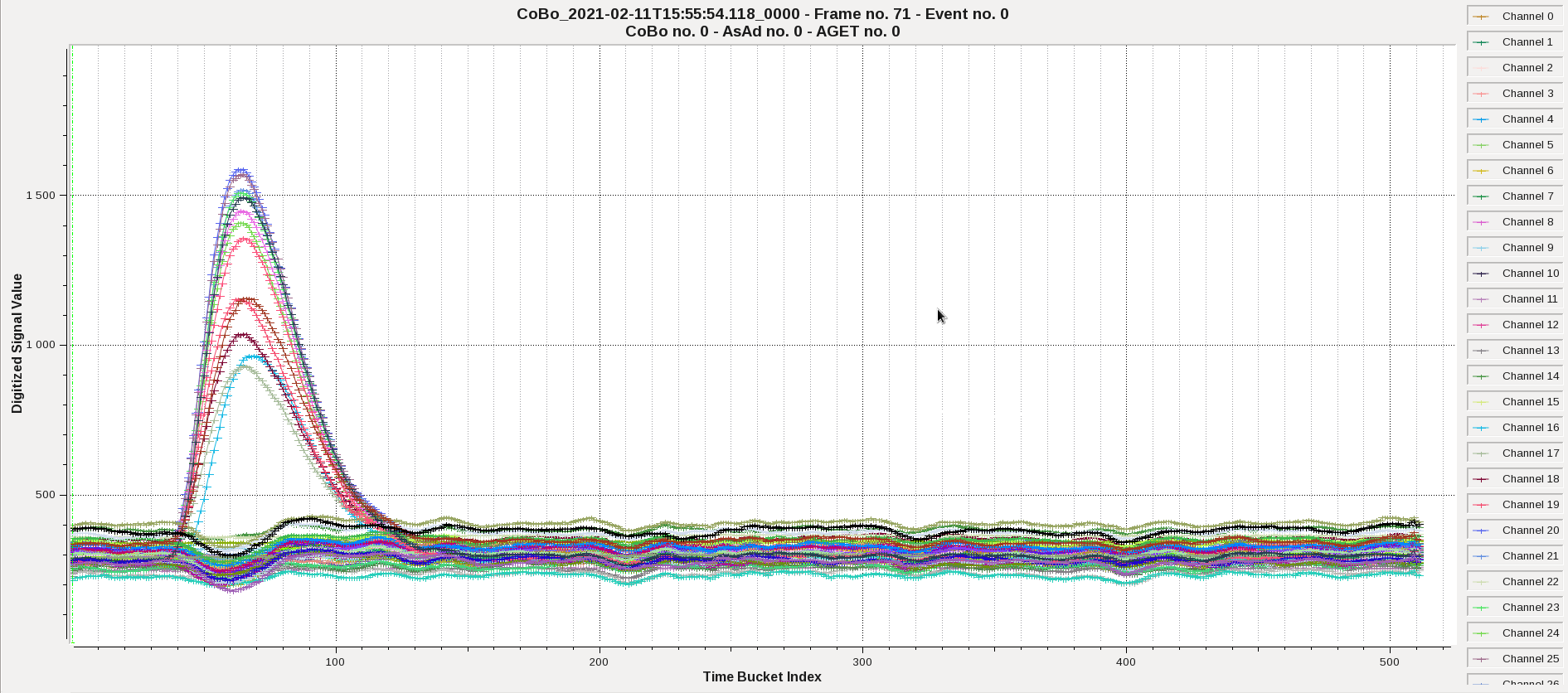}
	\caption{Example of a recorded spectrum obtained with GetController with the proton micro-beam}
	\label{Cobo-protons}
\end{figure}

The software used to control the acquisition and visualize data is GetController, developed by the CEA/IRFU, Saclay. Example of spectra obtained with the GetController acquisition system are available in figure \ref{Cobo-alpha} with an $\alpha$ source (see section~\ref{source}) and in figure~\ref{Cobo-protons} with a proton micro-beam (see section~\ref{micro-beam}). Each curve of the spectra represents the time evolution of a pad signal (10 ns per time bin).

\section{Detection efficiency measurements}
\label{efficiency}

With the proton recoil technique, the measured fission cross section is equal to the $^1$H(n,n)p cross section corrected by different ratios, as shown in equation (\ref{eq1}) \cite{Paola2017}: 

\begin{equation}
\sigma (n,f)(E_n) = \overline{\sigma_{H(n,n)p}} \frac{N_{ff}}{N_p} \frac{\varepsilon_p}{\varepsilon_F} \frac{N_H}{N_{target}} \frac{\Omega_{H-foil}}{\Omega_{target}} \label{eq1}
\end{equation}

Where:\\
- $\sigma (n,f)(E_n)$ is the measured fission cross section;\\
- $\overline{\sigma_{H(n,n)p}}$ is the elastic scattering cross section averaged on the energy range;\\
- $N_{ff}/N_p$ is the ratio of the detected quantities of fission fragments and of recoil protons;\\
- $\varepsilon_p/\varepsilon_F$ is the ratio of the total efficiencies of the protons detector and of the fission fragments detector;\\
- $N_H/N_{target}$ is the ratio of the hydrogen quantity in the H-rich foil and of nuclei present in the fissile target;\\
- $\Omega_{H-foil}/\Omega_{target}$ is the ratio of the solid angles between the H-rich foil and the target.\\

One can see in equation (\ref{eq1}) that the cross section accuracy is linearly dependent on the accuracy of the protons detector efficiency. As the elastic diffusion is isotropic at low neutron energies, the detection efficiency is the product of two factors (eq.~\ref{eq2}):
\begin{equation}
\varepsilon_{detection} = \varepsilon_{intrinsic} \times \varepsilon_{geometric} \label{eq2}
\end{equation}

When working with a source (see section~\ref{source}), the emission of particules is considered as isotropic and the geometric efficiency depends on the opening solid angles and is determined by a simple simulation. Under neutron irradiation, the proton recoil emission is also isotropic but the geometric efficiency must take kinematics effects into account.

The intrinsic efficiency is the ratio between the number of particles entering the detector and the number of particles actually detected. Some detectors, for example the silicon detectors, have an intrinsic efficiency of 100\% for charged particles. For other detectors, the intrinsic efficiency must be measured, which can be difficult. It often includes complex simulations and introduces additional uncertainties.

To have a detection efficiency as accurate as possible for the GPRT, the intrinsic efficiency must be 100\%: each proton entering the detector must be detected. Significant effort has been invested in verifying this point. Several different experiments were carried out to determine the detection efficiency, and thus subsequently determine the intrinsic efficiency. Firstly, an $\alpha$ source used, described in section~\ref{source}. Secondly, a mono-energetic proton micro-beam was used for more precise results, with varying proton energies and rates. These tests are described in section~\ref{micro-beam}. Finally, test bench measurements of the detector dead time were performed, and these are described in section~\ref{test_bench}.

\subsection{Tests with an $\alpha$ source}
\label{source}

The activity of the $\alpha$ source (composed of $^{239}$Pu, $^{241}$Am and $^{244}$Cm with primary $\alpha$ energies of 5.155~MeV, 5.486~MeV and 5.806~MeV respectively) was accurately measured using $\alpha$ spectroscopy: a silicon detector at a long distance (122.3~mm) over a long period of time (40~hours). A Monte Carlo simulation has been used to calculate the geometric efficiency of this setup. A sensitivity analysis of the parameters on the geometric efficiency has also been performed. The obtained activity of the source is 517.9~Bq~$\pm$~0.85\%$_{sys}$~$\pm$~0.37\%$_{stat}$ (more details, see \cite{ChatelPhD}, ch.~5.4.1.2).

This source was then placed on the macor sample disk, at the entrance of the GPRT, right before the first collimator. The same Monte Carlo simulation was used to accurately calculate the geometric efficiency of the GPRT. The parameters used with their errors, as well as the sensitivity of the geometric efficiency, are presented in Table \ref{tab:geom_eff}, from \cite{ChatelAnimma}. The simulated geometric efficiency of the GPRT device using an $\alpha$ source is (0.871~$\pm$~0.032)\%. 

\begin{table*}[h!]
\caption{\label{tab:geom_eff}Parameters to calculate the GPRT's geometric efficiency \cite{ChatelAnimma}}
\begin{tabular}{ c c c c }
\hline
Parameters & Values& Absolute uncertainties& Sensitivity\\
& (mm) & (mm) & (\%/mm) \\ \hline
Collimator radius & 5.00 & 0.01 & 38.91 \\ 
Source radius & 4.10 & 0.25 & 0.82 \\ 
Source-2$^{nd}$ collimator distance & 26.2 & 0.5 & 7.33 \\ 
\hline
\end{tabular}
\end{table*}

Acquisitions were then run with a gas pressure of 100~mbar. With this pressure, the bias voltages are of -430~V for the mesh and -2000~V for the cathode. Under these conditions, the energy of the $\alpha$ particles is too high to be stopped in the GPRT. However, the energy deposition (below 100 keV/cm) is similar to what one could have with recoil protons. To be considered as a good event, the particle must pass through both collimators before being detected in the E chamber.

To find the optimal parameters for the intrinsic efficiency and to reject as many parasitic events as possible, tests were performed with several gains and amplitude thresholds. The detection efficiency was then measured. Considering the activity of the $\alpha$ source and the geometric efficiency, the expected $\alpha$ detection rate was of 4.5~$\alpha$/s. Measuring the detection rate with optimal conditions, the obtained intrinsic efficiency is: (99.2~$\pm$~3.7$_{sys}$~$\pm$ 1.1$_{stat}$)\%. Tests proved that the threshold does not influence the efficiency for such high-energy particles. 

This efficiency is in agreement with the 100\% one could have expected. However, the large systematic uncertainty prevents from confirming the 100\%. Most of this uncertainty comes from the geometric efficiency. One should note here the very limiting factor that will be present in most experiments: the distance between the source, placed where the H-rich foils should be placed in an experiment, and the collimator is not well enough known due to geometric constraints (see table \ref{tab:geom_eff}, last line). This will have to be improved when designing the final version of the GPRT.

To obtain a better evaluation of the intrinsic efficiency in the detection of low energy protons, which is the main goal of the GPRT, tests have then been carried out with a proton beam.

\subsection{Tests with a proton micro-beam}
\label{micro-beam}

\subsubsection{Setup description}
\label{micro-beam_setup}

The aim of this experiment was to directly send a proton beam of an accelerator into the GPRT.

During a cross section measurement, the expected rate of recoil protons is of the order of 1 proton/s. To be as close as possible to those conditions and to perform an accurate efficiency measurement, a micro-beam of direct protons was used. Moreover, with a mono-directional beam, a controlled trajectory is assured, implying no geometric efficiency to be considered. The proton flux can be monitored by a silicon detector placed downstream and the rate of the direct proton beam can vary to study the GPRT efficiency as a function of this rate. 

All those requirements have been fulfilled by the beam delivered at the AIFIRA facility located on the LP2i site, a light ion beam facility for ion beam analysis and irradiation \cite{AIFIRA}. This accelerator is a Singletron able to deliver protons, deuterons and helium ions in the MeV energy range. Its maximum voltage is 3.5~MV, its maximum proton flux is 10$^{13}$ protons/s, with a usual one of 10$^{10}$ protons/s for micro-irradiation. The energy precision of the delivered protons is 10$^{-5}$ with an excellent stability over time. This accelerator contains five different beam lines. The only one able to accommodate the significant dimension of the setup is the physics line. It is dedicated to the production of mono-energetic neutrons and detector characterization but was, however, not yet used to produce micro-beams.

\begin{figure}[h!]
	\centering
\begin{subfigure}{0.5\linewidth}
	\centering
	\includegraphics[width=1\linewidth]{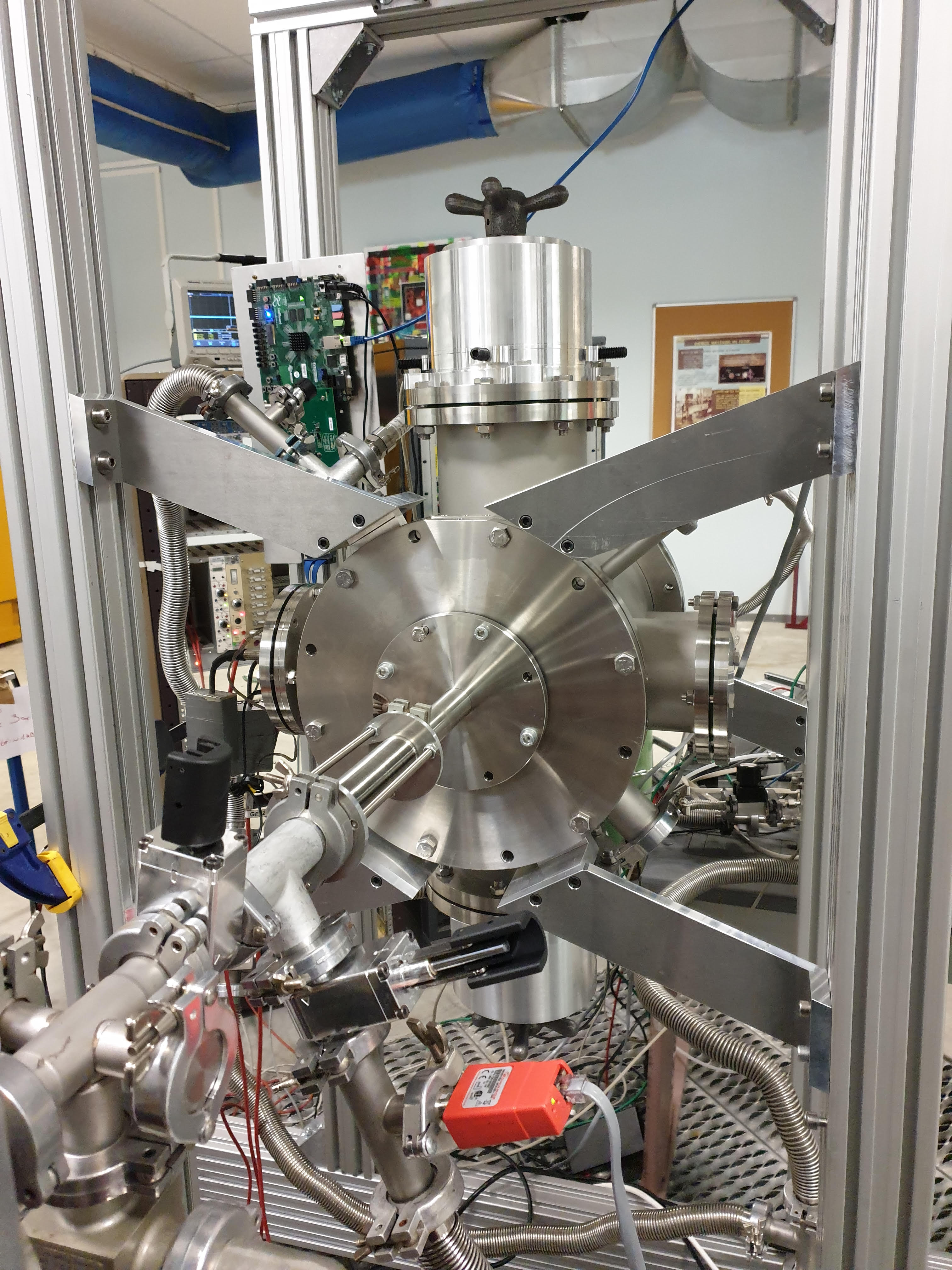}
	\caption{\label{Chambre_photo}}
\end{subfigure}\hfill
\begin{subfigure}{0.5\linewidth}
	\centering
	\includegraphics[width=0.98\linewidth]{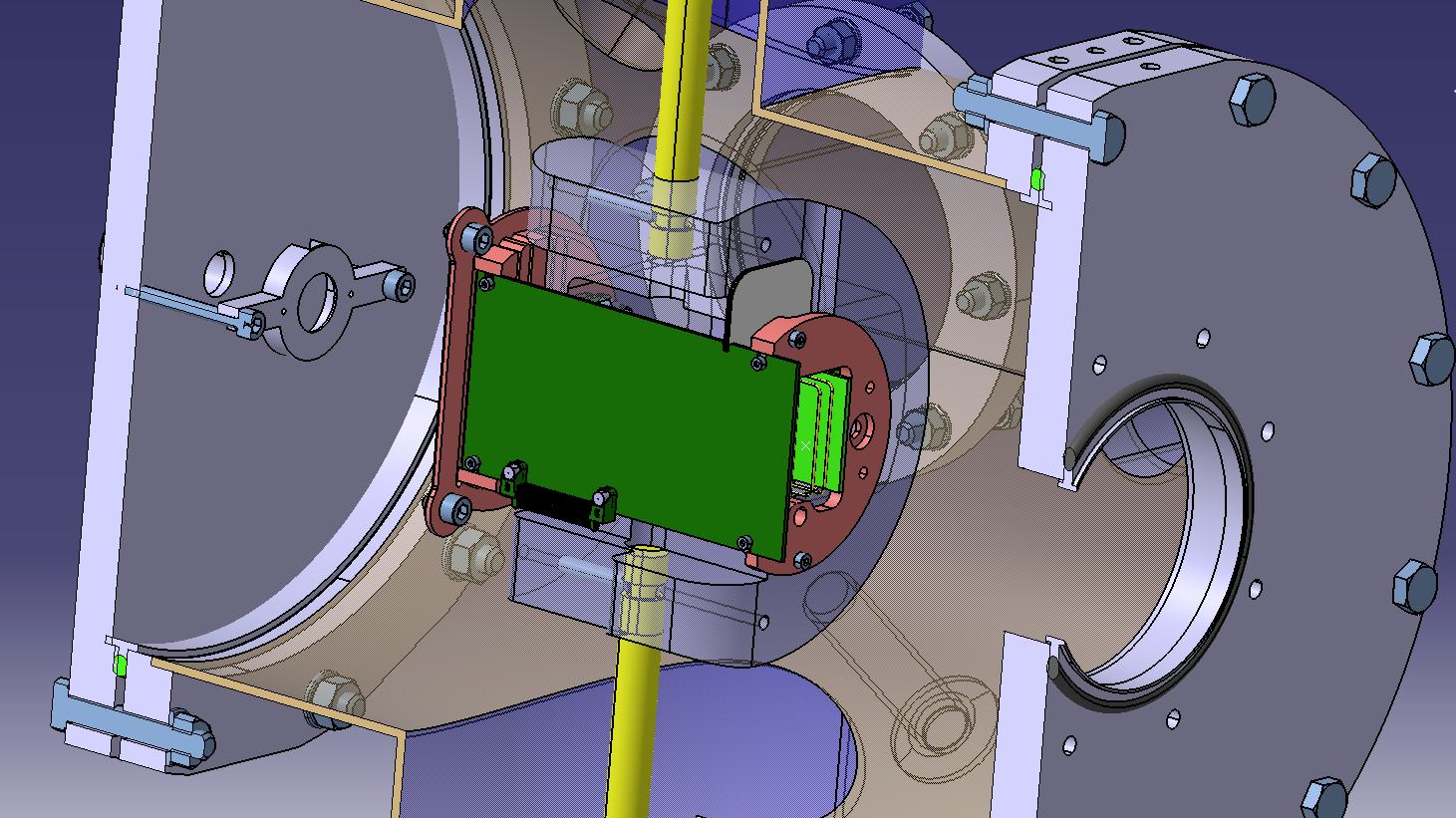}
	\caption{\label{Chambre_scheme}}
\end{subfigure}
	\caption{\label{Chambre_microbeam} Picture (a) and scheme (b) of the chamber designed for this experiment.}
\end{figure}

To send a proton micro-beam into the GPRT, a dedicated chamber, presented in figure \ref{Chambre_microbeam}, has been designed. The required characteristics were to be connectable to the accelerator beam-line, to be able to be instrumented and to allow the GPRT to move in two directions inside the chamber, so the proton-beam can be sent in different parts of the GPRT \cite{ChatelND}.

To obtain a micro-beam with an intensity as low as possible, the accelerator had to provide a beam with the minimum current and as much defocused as possible. The minimum current that can be measured by the accelerator was of 15~pA, which is still many orders of magnitude above the desired current. A first insulating collimator of 2~mm was placed at the end of the beam line to help centering the beam. A second one with a diameter of 20~$\mu$m was placed right behind the first one. The reading of the accelerator current was done from the second collimator. To separate the vacuum of the accelerator from the gas-filled chamber, an aluminum mylar window was placed after the second collimator. As protons can be scattered by the collimators and the window, a last collimator of 2~mm was placed at the entrance of the GPRT to prevent the scattered protons from entering the detector. Finally, a silicon detector was placed downstream the GPRT. Considered as having 100\% of intrinsic efficiency, this detector was used as reference to monitor the proton flux and then to measure the GPRT’s intrinsic efficiency. 

While developing this chamber, some extra features have been added. For example auxiliary holes have been drilled. The chamber can thus be used to characterize other detectors or for other applications.

\subsubsection{Intrinsic efficiency}
\label{micro-beam_efficiency}

The chosen energy of the protons was 3~MeV, this energy making it easier defocusing as much as possible the beam. The gas pressure chosen is still 100 mbar to minimize proton scattering in the gas. The power supplies of the cathode and of the electric mesh are identical to those used for the tests with $\alpha$-particles. The used acquisition parameters are the ones identified as giving the optimum conditions. Some tests have also been performed at 200~mbar to check the consistency of the results, with a mesh voltage set at -500~V.

Protons of 3~MeV are expected to deposit 14~keV/cm for 100~mbar and 30~keV/cm for 200~mbar. For comparison, 500~keV recoil protons at 80~mbar pressure (adapted to reach the Bragg peak before the end of the detector) deposit 45~keV/cm and 80 keV/cm at the Bragg peak. Therefore, the energies deposited by direct protons in this experiment are much lower than those deposited by recoil protons in the energy range of interest. However, as there is no neutron production, there is no generation of $\gamma$-rays nor electrons and very little electromagnetic parasites. The signals obtained during this experiment are very clear, with peak amplitudes well above the small baseline fluctuations. Those fluctuations are much weaker than the ones obtained with recoil proton experiments previously carried out \cite{Paola2018}. One must therefore remain cautious about the ability of GPRT to detect low deposited energies in parasited environments.

In this experiment, the proton trajectory has been chosen to be straight and well-defined to be sure that every proton passing through the GPRT is detected into the silicon detector. The efficiency has been measured by comparing the number of protons detected by the GPRT and by the silicon detector for a rate ranging from 0.6 protons/s to 350 protons/s.

The results are analyzed to exclude the rare parasitic events, of the order of 1\%. For rates below 3 protons/s, the intrinsic efficiency is, as expected, 100\%, which is consistent with the efficiency measured with the low activity $\alpha$ source. However, it decreases progressively down to 7\% at the maximum rate tested. The efficiency of 50\% is obtained for 38~protons/s, corresponding to an average effective dead time of 26 ms per event, much higher than the theoretical 1.4 ms given in section \ref{Acquisition}. 

The GPRT has then been tested with the “external trigger” mode, by triggering with the silicon detector, to investigate if the trigger process was the source of the issue. Results with this external trigger are in agreement with the previous ones.

These results cast doubts on the acquisition system itself. To investigate this hypothesis, tests were carried out on a test bench.

\subsection{Results with a test bench}
\label{test_bench}

The GPRT electronic Zedboard has first been connected to a test bench which sends electrical pulses at a given frequency. Tests were performed at frequencies from 10 Hz to 1000 Hz, with internal and external triggers. With a fix dead time and a regular frequency, one can expect to go directly from 100\% down to 50\%, when half the events cannot be processed because they are too close to the previous one.

However, the results give 100\% of efficiency for a frequency of 10 Hz and decrease progressively for higher frequencies. The average dead time is of 26~ms at 100~Hz and 38~ms at 1000~Hz with the internal trigger mode and of 19~ms for 100~Hz and 15~ms for 1000~Hz with the external trigger mode. Moreover, tests have shown a poor reproducibility of the results, depending on external parameters like acquisition reboot or computer reboot.

When analyzing the data, it has been noticed that some events were missing in bunches, and the quantity seemed to be random. This has been explained by the way the data were transferred to the computer: the data were stored in the acquisition card before being transferred to the computer. During the transfer process, some events were lost. This problem has been circumvented by transferring data after each event while waiting for the transfer error to be corrected. In this way, no more events were “lost” during the data transfer.

Tests were carried out with this patch. The efficiency was found to be 100\% up to the frequency of 137~Hz when it drops to 50\%. This corresponds to a dead time of 7.3~ms, which means 5.9~ms higher than what was expected in section \ref{Acquisition}. This difference of time is associated to the data transfer, much longer than the data read-out time. This data transfer time is still under investigation by the electronic service of the LP2i.

\subsection{Discussion}
\label{Discussion}

With a low count rate, the GPRT has an intrinsic efficiency of 100\%. However, the dead time is much higher than expected. If it were 1.4~ms as foreseen, 10~protons per second would give 1.4\% of dead time. With 7.3~ms, the dead time becomes 7.3\%, which is barely acceptable for a high precision measurement. However, for cross section measurements, only 1 to 2~protons per second are expected in the GPRT. With such low rates, the 7.3~ms of dead time still allows an accurate measurement of the neutron flux. Nevertheless, there are much more physical events happening in the GPRT; for instance all recoil protons stopped in the $\Delta$E chamber by the collimator. There could also be some parasitic events due to electromagnetic perturbations, which may happen in short burst. It is thus very important to have good trigger conditions but also to have some margin on the dead time.

Until a fix is found, a solution to reduce the dead time is to work in “partial readout” mode. In this case, only the data of activated pads will be sent to the computer. With usually between 10 to maximum 30 channels firing, this mode can allow a reduction of the dead time of a factor 2 to 3. However, with this solution, signals below threshold are not recorded. Another potential solution would be to reduce the recording from 512 points to 256. A reduction of the dead time of a factor 2 is then expected.

\section{Conclucion}
\label{conclusion}

The Gaseous Recoil Proton Telescope is currently developed at the LP2i. It is a functional prototype whose aim is to enable a precise neutron flux measurement between 200~keV and 2.5~MeV thanks to the recoil proton technique. This should allow accurate measurements of fission cross sections.

The intrinsic efficiency, required to be at 100\% for an accurate measurement, has been deeply investigated. It is at 100\% for a low rate of particles but drops significantly when the rate increases. This has been identified to be caused by a problem during data transfer process. For years, the GPRT has suffered from slow and deficient data transfer. The issue has now been identified and a patch has been implemented to ensure that events are no longer randomly lost during data transfer. A transfer time of 5.9~ms per event has also been identified and is under investigation. Although this dead time is higher than expected, the GPRT now meets the minimal requirements to be functional. One still needs to keep the counting rate low enough to ensure a 100\% intrinsic efficiency and a manageable dead time correction.

The GPRT is a prototype and a final design will be made, with a special care for the knowledge of the geometry, responsible of the main part of the systematic uncertainties as pointed out earlier in the text.

\section*{Acknowledgements}
The authors would like to thank the AIFIRA team work, always helping to prepare and during experiments. This project has received funding from the Euratom research and training program 2014–2018 under Grant No. 847552 (SANDA).

\end{document}